\documentclass[preprint,fleqn,showpacs,showkeys]{revtex4}
\usepackage{graphicx}
\usepackage{amssymb}
\usepackage{amsmath}
\usepackage{bm}
\usepackage{units}
\usepackage{mathtools}
\usepackage[T1]{fontenc} 
\usepackage[utf8]{inputenc}
\usepackage[english]{babel}
\usepackage{hyperref}
\usepackage{lineno}
\usepackage[section]{placeins}
\usepackage{subfig}
\usepackage{multirow}

\begin{document}
\setcounter{page}{0}
\title[]{Momentum-kick model application \\ to high multiplicity pp collisions at \unit[$\sqrt{s}=13$]{TeV} at the LHC}
\author{Beomkyu \surname{Kim}}
\author{Hanul \surname{Youn}}
\author{Beomkyu \surname{Kim}}
\author{Hanul \surname{Youn}}
\author{Soyeon \surname{Cho}}
\author{Jin-Hee \surname{Yoon}}
\affiliation{Department of Physics, Inha University, Republic of Korea}

\date[]{Received 16 April 2020}

\begin{abstract}
In this study, the momentum-kick model is used to understand the ridge behaviours in dihadron $\Delta\eta$--$\Delta\varphi$ correlations recently reported by the LHC in high-multiplicity proton-proton (pp) collisions. The kick stand model is based on a momentum kick by leading jets to partons in the medium close to the leading jets. The medium where partons move freely is assumed in the model regardless of collision systems. This helps us apply the method to small systems like pp collisions in a simple way. Also, the momentum transfer is purely kinematic and this provides us a strong way to approach the ridge behaviour analytically. There are already several results with this approach in high-energy heavy-ion collisions from the STAR and PHENIX at RHIC and from the CMS at LHC. The momentum-kick model is extended to the recent ridge results in high-multiplicity pp collisions with the ATLAS and CMS at LHC. The medium property in high-multiplicity pp collisions is diagnosed with the result of the model. 
\end{abstract}



\maketitle

\section{INTRODUCTION}

Mainly, a ridge behaviour in dihadron $\Delta\eta$--$\Delta\varphi$ correlations is a well known property of high-energy heavy-ion collisions, which originates from the collective behaviour of strongly interacting matter, so called the Quark-Gluon Plasma (QGP)~\cite{Chatrchyan:2011eka,CMS:2013bza}. Recently, the LHC reported the ridge behaviour in dihadron correlations in high-multiplicity proton--proton (pp) collisions at centre-of-mass energies $\sqrt{s}$ = 2.76 to \unit[13]{TeV}~\cite{Khachatryan:2010gv,Aad:2015gqa,Khachatryan:2015lva}. The observation is also successfully extended to those in p--Pb collisions at the LHC~\cite{Abelev:2012ola,Aad:2012gla,CMS:2012qk,Aad:2014lta}.  
However, the origin of the ridge behaviours in small systems like pp and p--Pb collisions is still controversial.

In relativistic central heavy-ion collisions, the ridge behaviour is understood as a result from hydrodynamic flows of the QGP matter~\cite{Xu:2011jm}. The same approach was applied to study the ridge result specially in high-multiplicity pp and the most central p--Pb collisions at the LHC energies~\cite{Werner:2010ss,Bozek:2012gr,Shuryak:2013ke,Kalaydzhyan:2015xba}.
However, these systems are too small to make a reliable size of a strongly interacting matter where the long-range correlation (ridge) takes place. 
There are several ideas to explain the long range correlations in small systems. These ideas are based on some constraints on Multiple Parton Interactions (MPIs) that is expected to be more pronounced in high multiplicity pp collisions~\cite{Sjostrand:1987su,Abelev:2013sqa,Abelev:2014mva}. The constraints include the colour reconnection (CR)~\cite{Christiansen:2015yqa,Ortiz:2013yxa} and hydrodynamic approach to partons in MPIs~\cite{Pierog:2013ria,Drescher:2000ha}.

The momentum kick stand model~\cite{Wong:2007pz,Wong:2008yh,Wong:2009cx,Wong:2011qr} assumes that the leading jets in high multiplicity pp collisions collide with the closest partons in the medium constructed by the MPIs and then a fractional momentum transfer happens to the partons in the direction of the jets. The methodology of the model with the momentum transfer is purely kinematic compared to other models and provide a simple but strong way to explain the ridge behaviour in small systems like a pp collision. The model has been validated for measurements at the RHIC and LHC from pp to heavy-ion collisions~\cite{Wong:2007pz,Wong:2008yh,Wong:2009cx,Wong:2011qr,Youn:2017beo}.

This document addresses the details of the kick stand model in Sec.~\ref{sec:modeldescription} and the result of the model with the parameters in pp high multiplicity conditions are applied to the recent results at ATLAS and CMS in Sec.~\ref{sec:analysis}. Then, conclusions are followed in Sec.~\ref{sec:conclusion}.  

\section{The kick stand model}
\label{sec:modeldescription}

\begin{figure}[!htb]
\includegraphics[width=0.7\textwidth]{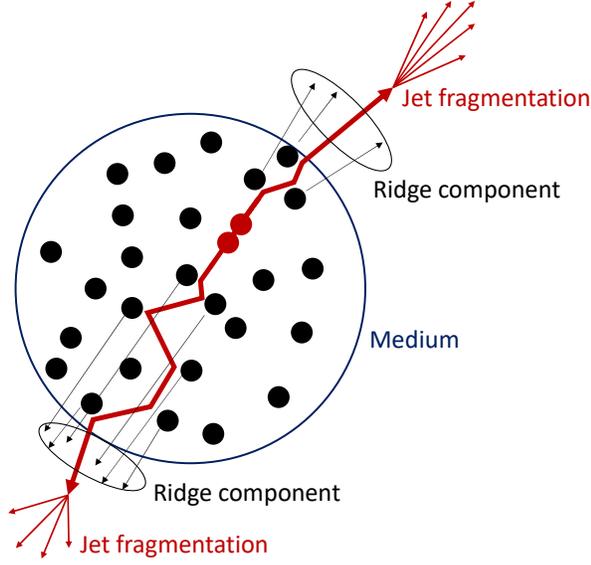}
\caption{(Color online) The leading $2\rightarrow2$ hard scattering that is represented as jets (red solid lines) brings a momentum kick to nearby partons along the jet direction that are finally constructing the ridge components (black solid line). }
\label{fig:kickstandmodel}
\end{figure}

The kick stand model basically assumes a medium with partons that can be created by the MPIs, especially in high-multiplicity pp collisions. As described schematically in Fig.~\ref{fig:kickstandmodel}, the leading hard partons (jets) pass through the medium by interacting with nearby partons and finally hadronize to jet constituent particles. The nearby partons in the medium acquire a momentum kick $\vec{q}$ along the jet direction and construct a cluster, which hadronizes to final stable particles.

The final transverse momentum ($p_{\mathrm{T}_f}$) distribution can be thought as the combination of the momentum distributions of the ridge and jet fragment as (see also Refs.~\cite{Wong:2007pz,Wong:2008yh,Wong:2009cx,Wong:2011qr,Youn:2017beo})
\begin{align}
    \left[\frac{1}{N_\mathrm{trig}}\frac{\mathrm{d} N_\mathrm{ch} }{p_{\mathrm{T}_f} \mathrm{d} p_{\mathrm{T}_f} \mathrm{d}\Delta \eta \mathrm{d} \Delta \varphi}\right]_\mathrm{total} &  = f_R\frac{2}{3}\langle N_k\rangle\left[\frac{\mathrm{d} N }{p_{\mathrm{T}_f} \mathrm{d} p_{\mathrm{T}_f} \mathrm{d}\Delta \eta \mathrm{d} \Delta \varphi}\right]_\mathrm{ridge} \nonumber \\ & \quad \quad \quad \quad \quad \quad +f_J\left[\frac{\mathrm{d} N_\mathrm{jet}}{p_{\mathrm{T}_f} \mathrm{d} p_{\mathrm{T}_f} \mathrm{d}\Delta \eta \mathrm{d} \Delta \varphi}\right]_\mathrm{jet} \quad,
   \label{eq:totaldistribuiton}
\end{align}
where $N_\mathrm{ch}$ is the number of charged particles, $N_\mathrm{jet}$ is the total number of near-side (charged) jet fragments associated with the $p_\mathrm{T}$ trigger, $f_R$ and $f_J$ are the survival factors of ridge particles and the jet fragments, respectively, that can be detected experimentally in the detector, $\langle N_k \rangle$ is the average number of kicked medium parton per jet and the factor 2/3 represents the portion of charged particles for the associated particles, and $\Delta \eta = \eta - \eta_\mathrm{jet}$ and $\Delta \varphi = \varphi-\varphi_\mathrm{jet}$ represent the differences in the pseudorapidities and azimuthal angles between a jet and associated particle, respectively. 

When the initial parton momentum distribution $P_i(\vec{p_i})$ is modified to $P_f(\vec{p_f})$ by the momentum kick $\vec{q}$, the ridge part can be described as 
\begin{equation} 
\label{eq:momentumkick}
\frac{\mathrm{d} N}{p_{\mathrm{T}_f} \mathrm{d} p_{\mathrm{T}_f} \mathrm{d}\Delta \eta \mathrm{d} \Delta \varphi}  = \left[  \frac{\mathrm{d} N_i }{p_{\mathrm{T}_i} \mathrm{d} p_{\mathrm{T}_i} \mathrm{d}\Delta y_i \mathrm{d} \Delta \varphi_i } \right]_{p_i=p_f-q} \times \sqrt{1-\frac{ m_\pi^2 }{m^2_{{\mathrm T}_f}} \cosh^2{y_f}} \quad, 
\end{equation}
where $m_\pi$ and $m_\mathrm{T} = \sqrt{m_\pi^2 + p_\mathrm{T}^2}$ are the rest and transverse masses of partons  assumed as pions, and where $y$ and $\eta$ are the rapidity and pseudorapidity of partons, respectively. In the equation, the first and second terms on the right side represent the initial parton momentum distribution and the Jacobian from rapidity to pseudorapidity phase space, respectively. It is worth noting that the initial parton momentum distribution can be expressed in terms of the temperature ($T$) of the medium that can be obtained from the experimental data as given by equation
\begin{equation}
    \quad \quad \quad \quad \quad \frac{\mathrm{d} N_i }{p_{\mathrm{T}_i} \mathrm{d} p_{\mathrm{T}_i} \mathrm{d}\Delta y_i \mathrm{d} \Delta \varphi_i } = A_\mathrm{ridge} (1-x)^a \frac{\exp{(-m_{\mathrm{T}_i}/T)}}{\sqrt{m_\pi^2+p_{\mathrm{T}_i}^2}} \quad,
\end{equation}
where $A_\mathrm{ridge}$ is the normalization constant and $x = (p_{0_i} + p_{3_i})/(p_{0_\mathrm{beam}} + p_{3_\mathrm{beam}})$ is the light-cone variable indicating the ratio of the transverse momenta between a medium parton ($i$) and beam particle (beam).

The jet part is expressed as
\begin{align}
    \frac{\mathrm{d} N_\mathrm{jet}}{p_{\mathrm{T}_f} \mathrm{d} p_{\mathrm{T}_f} \mathrm{d}\Delta \eta \mathrm{d} \Delta \varphi}  & = N_\mathrm{jet} \frac{\exp\{(m_\pi-m_{T_f})/T_\mathrm{jet}\}}{T_\mathrm{jet} (m_\pi + T_\mathrm{jet})} \nonumber \\ & \quad \quad \quad \quad \quad \quad \quad \quad \quad \times \frac{1}{2\pi \sigma^2_\varphi} \exp{\{-\left[ (\Delta \varphi)^2 + (\Delta \eta)^2 \right] / 2\sigma^2_\varphi  \} }\quad.
\end{align}
See also for the detailed description in Refs. \cite{Wong:2007pz,Wong:2008yh,Wong:2009cx,Wong:2011qr,Youn:2017beo}.     

\section{Analysis}
\label{sec:analysis}

The momentum-kick model is applied to the ATLAS and CMS results in pp collisions at centre-of-mass energy $\sqrt{s}$~=~\unit[13]{TeV}~\cite{Aad:2015gqa,Khachatryan:2015lva}. 
The ridge yield at ATLAS is presented for the transverse momentum of both trigger and associated particles in $0.5<p_\mathrm{T}^\mathrm{trig},~p_\mathrm{T}^\mathrm{asso}<5\,\mathrm{GeV}/c$ for events with the number of offline charged particles $N_\mathrm{ch}^\mathrm{offline}$ measured in $|\eta|<2.5$ above 120. The CMS provides the result for $1<p_\mathrm{T}^\mathrm{trig},~p_\mathrm{T}^\mathrm{asso}<4\,\mathrm{GeV}/c$ and $N_\mathrm{ch}^\mathrm{offline}>135$.

To avoid the jet contribution in the yield of the near-side ridge, the ATLAS and CMS applied a large $\Delta \eta$ cut ($|\Delta \eta|>2$) for pairs of trigger and associated particles in dihadron correlations. It is worth noting that the ridge yield relies on a correlated portion between jets in the near side at $\Delta \varphi \sim 0$ and  dijets in the away side at $\Delta \varphi \sim \pi$. The ridge yield at ATLAS was corrected for the correlated portion by using the template method~\cite{Aad:2015gqa}, on the other hand, the CMS corrected the dijet correlated yield using the traditional ZYAM hypothesis~\cite{Ajitanand:2005jj}. 

\begin{table}[hbt!]
\begin{tabular}{ c|c|c|c|c } 
\hline
& \quad \multirow{2}{*}{PHENIX (Au--Au)} \quad & \quad \multirow{2}{*}{CMS (Pb--Pb)} \quad & \quad CMS (pp) \quad & \quad ATLAS, CMS (pp) \quad \\
&  &  & Min. Bias \quad & \quad High Mult. \quad \\
\hline
$\sqrt{s}$, $\sqrt{s_\mathrm{NN}}$ (TeV) & 0.2 & 2.76 & 7 & 13 \\
\hline
$T$ (GeV) & 0.5 &  0.6  & 0.7 & 1.54 \\
$T/T_{\rm PHENIX}$ & -- &  1.2  & 1.4 & 3.08 \\

\hline
\end{tabular}
\caption{The temperature $T$ of the medium dependent on centre-of-mass energy is increased with respect to the RHIC case at $\sqrt{s_\mathrm{NN}}=200\,\mathrm{GeV}$ ($T/T_{\rm PHENIX}$) by the method and information described in Refs.~\cite{Wong:2008yh,Wong:2011qr,Chatrchyan:2011av,Aad:2016mok}. Note that the parameters in pp collisions at $\sqrt{s} = 7\,\mathrm{TeV}$ for the CMS result is set to the minimum bias events (Min. Bias)~\cite{Khachatryan:2010gv}, while those are for high multiplicity events (High Mult.) at $\sqrt{s} = 13\,\mathrm{TeV}$ for the ATLAS and CMS results~\cite{Aad:2015gqa,Khachatryan:2015lva}. } 
\label{tab:scaleup}
\end{table}

The physical parameters that are used for the ATLAS and CMS results are reported in Tab.~\ref{tab:scaleup}. The parameters are calculated based on the RHIC result~\cite{Wong:2008yh,Adams:2004pa,Adams:2005ph}.
The temperatures of the medium ($T$) 
are dependent on centre-of-mass energy and those are increased accordingly. The medium temperature in pp collisions at $\sqrt{s} = 13\,\mathrm{TeV}$ is increased by about 300\% with respect to the case of the RHIC in Au--Au collisions at $\sqrt{s_\mathrm{NN}}=200\,\mathrm{GeV}/c$ following the description in Ref.~\cite{Wong:2011qr}. To determine the temperature of the medium, the $p_\mathrm{T}$ distribution of inclusive charged particles is used~\cite{Chatrchyan:2011av,Aad:2016mok}. 

\begin{table}[hbt!]
\begin{tabular}{@{} c|c|c|c|c|c|c|c @{}} 
\hline
&  \multirow{2}{*}{PHENIX (Au--Au)}  &  \multirow{2}{*}{CMS (Pb--Pb)}  &  \, CMS (pp) \,  &  \multicolumn{2}{c|}{ ATLAS (pp) }   & \multicolumn{2}{c}{ CMS (pp) } \\
&  &  & \, Min. Bias \, &  \multicolumn{2}{c|}{High Mult.}  & \multicolumn{2}{c}{High Mult.}\\

\hline
$\sqrt{s}$, $\sqrt{s_\mathrm{NN}}$ (TeV) & 0.2 & 2.76 &7 & \multicolumn{2}{c|}{13} & \multicolumn{2}{c}{13}\\
\hline
$q$ (GeV/$c$) & 0.8 & 0.7 & 2 & $\quad2\quad$ & $\quad1.2\quad$ & $\quad2\quad$ & $\quad1.2\quad$ \\
$f_R \langle N_k\rangle$ & 3 & 2.06--6.82  & 1.5 & 0.7 & 1 & 0.54 & 0.74\\
\hline
\end{tabular}
\caption{ Physical parameters of the momentum kick model. The results of PHENIX in Au--Au collisions at $\sqrt{s_\mathrm{NN}}$~=~\unit[200]{GeV} and CMS in Pb--Pb collisions at $\sqrt{s_\mathrm{NN}}$ = \unit[2.76]{TeV} and pp collisions at $\sqrt{s}$ = \unit[7]{TeV} are quoted from Refs.~\cite{Wong:2009cx,Wong:2011qr,Youn:2017beo}. The resulting parameter $q=1.2\,\mathrm{GeV}/c$ for the ATLAS and CMS data at $\sqrt{s}$ = \unit[13]{TeV}~\cite{Aad:2015gqa,Khachatryan:2015lva} is the fitting results of this study. Note that $q=2\,\mathrm{GeV}/c$ at $\sqrt{s} = 13\,\mathrm{TeV}$ is the fixed value (input value for the fit) for comparison to the result at $\sqrt{s}=7\,\mathrm{TeV}$. }
\label{tab:fittingresult}
\end{table}

\begin{figure}[hbt!]
 
\subfloat[]{%
  \includegraphics[width=0.48\columnwidth]{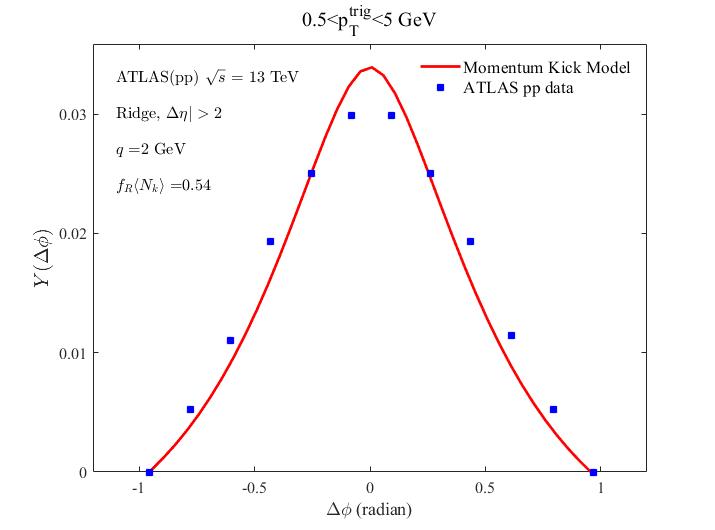}%
}
\subfloat[]{%
  \includegraphics[width=0.48\columnwidth]{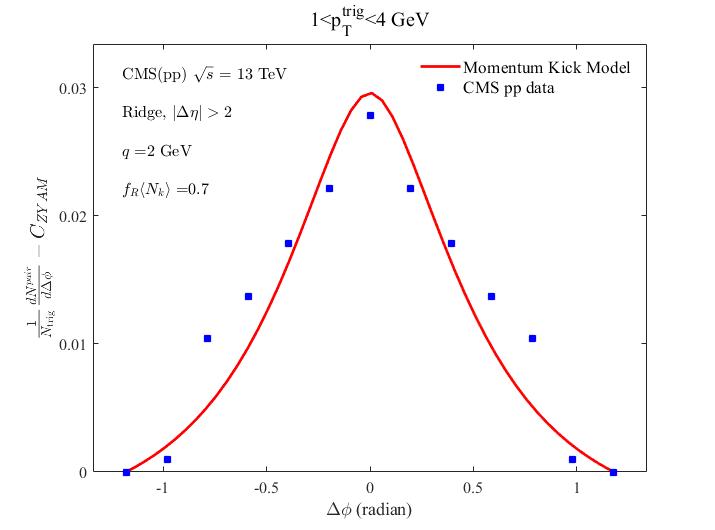}%
}

 \caption{ The momentum-kick model is fitted and compared to the ATLAS (a) and CMS (b) data for a momentum kick $q$ = \unit[2]{GeV/$c$}~\cite{Aad:2015gqa,Khachatryan:2015lva}.}
  \label{fig:mkmodelq20}
\end{figure}
\begin{figure}[hbt!]

\subfloat[]{%
  \includegraphics[width=0.48\columnwidth]{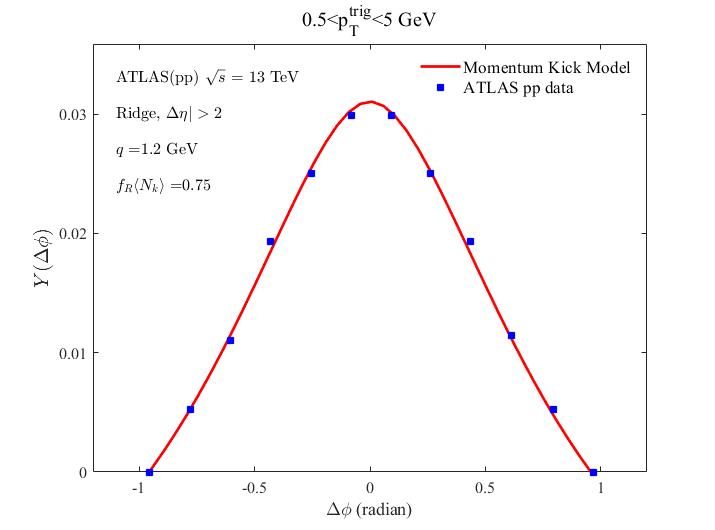}%
}
\subfloat[]{%
  \includegraphics[width=0.48\columnwidth]{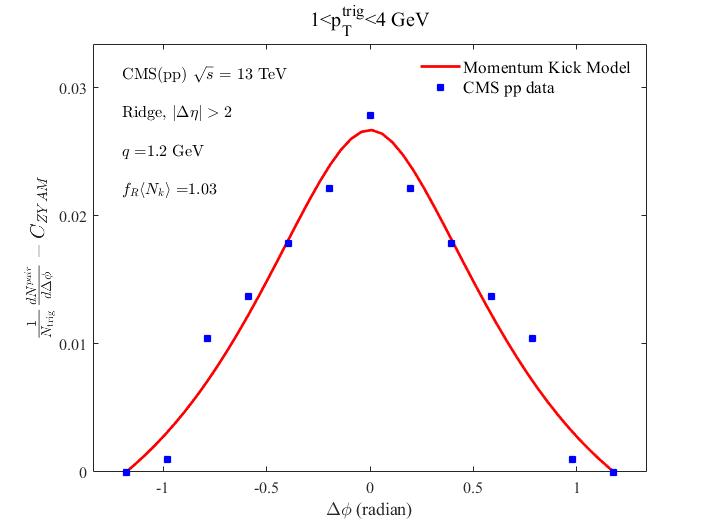}%
}

 \caption{ The momentum-kick model is fitted and compared to the ATLAS (a) and CMS (b) data for a momentum kick $q$ = \unit[1.2]{GeV/$c$}~\cite{Aad:2015gqa,Khachatryan:2015lva}.}
  \label{fig:mkmodelq12}
\end{figure}

The resulting fit values are listed in Tab.~\ref{tab:fittingresult}. The values of $q$ and $f_R \langle N_k\rangle$ are the fitting parameters that are determined with the experimental data. 
The temperatures of trigger jets ($T_\mathrm{jet}$) were fitted as $0.19+0.06 \times \langle p_\mathrm{T}^\mathrm{trig} \rangle$, $0.23+0.07 \times \langle p_\mathrm{T}^\mathrm{trig}\rangle$ and $0.27+0.08 \times \langle p_\mathrm{T}^\mathrm{trig} \rangle $ for PHENIX in Au--Au at $\sqrt{s_\mathrm{NN}}$ = 0.2~\cite{Wong:2009cx}, CMS in Pb--Pb at 2.76~\cite{Youn:2017beo} and CMS in pp collisions at \unit[7]{TeV}~\cite{Wong:2011qr}, respectively. 
The last two columns represent the results from the recent high multiplicity pp data at $\sqrt{s}=13\,\mathrm{TeV}$. Firstly, we tried $q=2\,\mathrm{GeV}/c$ as an input to compare the result to that from minimum bias events in pp collisions at $\sqrt{s}=7\,\mathrm{TeV}$.
At last, we find the best fit value of $q = 1.2\,\mathrm{GeV}/c$ for the high multiplicity pp data.
The fit results with the momentum kick $q = 2$ and \unit[1.2]{GeV/$c$} for the ridge yield as a function of $\Delta \varphi$ in dihadron correlations at LHC are shown in Fig.~\ref{fig:mkmodelq20} and Fig.~\ref{fig:mkmodelq12}, respectively. For both figures, the left sub-figures show for the ATLAS and the right ones for the CMS data. 

In the previous study~\cite{Wong:2011qr} \unit[$q=2$]{GeV/$c$} was used to fit the CMS data in minimum-bias pp collisions at $\sqrt{s}$ = \unit[7]{TeV}~\cite{Khachatryan:2010gv}. The new fit value $q=1.2$ describes the LHC data in high-multiplicity pp collisions at $\sqrt{s}$ = \unit[13]{TeV} well than \unit[$q=2$]{GeV/$c$} as compared in Fig.~\ref{fig:mkmodelq20} and \ref{fig:mkmodelq12}. 
This is interpreted that the average momentum kick $\langle q \rangle$ is decreased as more collisions occur per trigger jet in a denser medium that is prone to higher centre-of-mass energies. The same interpretation can be extended to the results in Au--Au and Pb--Pb collisions~\cite{Wong:2009cx,Youn:2017beo} as shown in Tab.~\ref{tab:fittingresult}. 
We can notice that the value of $f_R\langle N_k \rangle$ is increased as $q$ is decreased. As more momentum kicks are transferred from jets to partons, more partons can survive to detectors. In the same sense if you compare CMS (Pb--Pb) to CMS (pp) results, more partons are kicked in the medium produced in Pb--Pb collisions, and therefore $f_R\langle N_k \rangle$ becomes much larger. 

\begin{figure}[hbt!]

\subfloat[]{%
  \includegraphics[width=0.48\columnwidth]{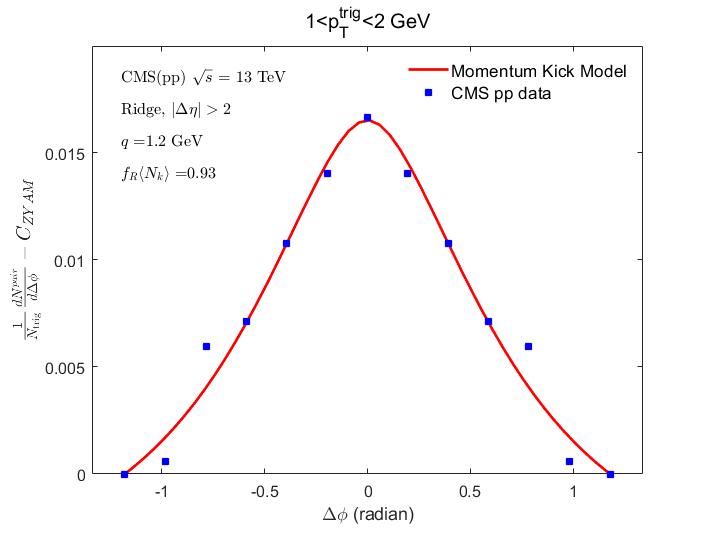}%
}
\subfloat[]{%
  \includegraphics[width=0.48\columnwidth]{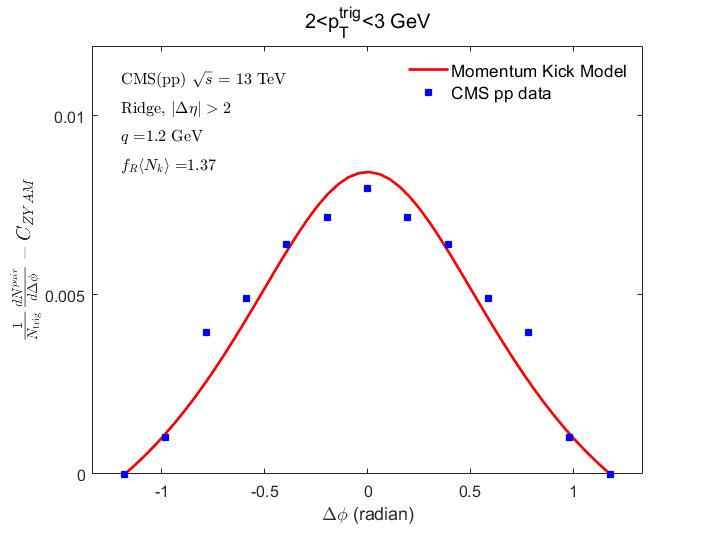}%
}
 
 \caption{ The momentum-kick model is applied to the ridge result of the CMS data in $1<p_\mathrm{T}^\mathrm{trig}<2\,\mathrm{GeV}$ (a) and $2<p_\mathrm{T}^\mathrm{trig}<3\,\mathrm{GeV}$ (b), separately, for a momentum kick $q$ = \unit[1.2]{GeV/$c$}~\cite{Aad:2015gqa}.}
  \label{fig:mkmodelq12cmspt}
\end{figure}

Figure~\ref{fig:mkmodelq12cmspt} shows fit results when the model is applied to the CMS data in two different $p_\mathrm{T}^\mathrm{trig}$ ranges, $1<p_\mathrm{T}^\mathrm{trig}<2\,\mathrm{GeV}$ (a) and $2<p_\mathrm{T}^\mathrm{trig}<3\,\mathrm{GeV}$ (b), with $q$ = \unit[1.2]{GeV/$c$}. 
When the value of $p_\mathrm{T}^\mathrm{trig}$ becomes higher from $1<p_\mathrm{T}^\mathrm{trig}<2$ to $2<p_\mathrm{T}^\mathrm{trig}<3\,\mathrm{GeV}$, an increased value of $f_R \langle N_k\rangle$ from 0.93 to 1.37 is observed, which implies that more kicked medium partons are survived and measured as a final state particle in the detector for higher $p_\mathrm{T}^\mathrm{trig}$ range. The measurement proves also the reliability of the kick stand model since this is an expected behaviour of the model.    

\section{Conclusion}
\label{sec:conclusion}
The framework of the momentum-kick model is applied to the near-side ridge yields in dihadron $\Delta \eta$--$\Delta \varphi$ correlations measured by ATLAS and CMS in pp collisions at $\sqrt{s}$ = \unit[13]{TeV}. The kick-stand model in pp collisions is based on the assumption that leading jets generated by the hard scattering provide a momentum kick $q$ to nearby partons in the medium created by the Multiple Parton Interactions in high-multiplicity proton--proton collisions. The model explains the ridge behaviour in dihadron correlations by the nearby partons boosted towards the direction of the leading jets.  

In the previous kick-stand model study~\cite{Wong:2011qr}, the CMS data in minimum-bias pp collisions at $\sqrt{s}$ = \unit[7]{TeV}~\cite{Khachatryan:2010gv} is described well with a momentum kick $q$ = \unit[2]{GeV/$c$}. The new fit value $q=1.2\,\mathrm{GeV}/c$ is extracted for the ATLAS and CMS data in high-multiplicity pp collisions at $\sqrt{s}$ = \unit[13]{TeV} in this study. The decreased momentum kick at $\sqrt{s}$ = \unit[13]{TeV} compared to the result at \unit[7]{TeV} can be understood that a denser medium is created for higher centre-of-mass energies because the average momentum kick $\langle q \rangle$ is expected to decrease for a denser medium due to increased partonic collisions as confirmed previously in Au--Au and Pb--Pb collisions~\cite{Wong:2009cx,Youn:2017beo}. We also observed that an increased value of $f_R \langle N_k\rangle$ is obtained as the value of $p_\mathrm{T}^\mathrm{trig}$ becomes higher in high-multiplicity pp collisions which implies that more kicked medium partons are survived and measured as a final state particle in the detector for higher $p_\mathrm{T}^\mathrm{trig}$. The measurement verifies the reliability of the kick stand model because the effect is an expected behavior of the model.

\begin{acknowledgments}

This work was supported by Inha University and the National Research Foundation of Korea (NRF) grant funded by the Korea government (MSIT) [No. NRF-2008-00458].

\end{acknowledgments}

\newpage
\bibliographystyle{utphys}   
\bibliography{biblio}

\end{document}